\documentclass[12pt]{article}
\usepackage{epsfig}
\usepackage{amsmath,amsfonts,amssymb}

\newcommand{\be}{\begin{equation}}
\newcommand{\ee}{\end{equation}}
\newcommand{\bem}{\begin{displaymath}}
\newcommand{\eem}{\end{displaymath}}
\newcommand{\ba}{\begin{eqnarray}}
\newcommand{\se}{\setcounter{equation}{0}}
\newcommand{\ea}{\end{eqnarray}}
\newcommand{\re}[1]{(\ref{#1})}

\newcommand{\1}{^{-1}}

\newcommand{\C}{{\cal C}}
\newcommand{\delt}{\delta^{\hspace*{-0.2mm}\mbox{\tiny G}}}
\newcommand{\dg}{^{\dagger}}
\newcommand{\di}{\mbox{d}\,}
\newcommand{\e}{\mbox{e}}

\newcommand{\f}{{\mbox{\scriptsize f}}} 
\newcommand{\g}{{\mbox{\scriptsize g}}} 
 
\newcommand{\bG}{\bar{G}} 
\newcommand{\G}{{\cal B}} 
\newcommand{\ga}{\gamma_5}
\newcommand{\h}{\frac{1}{2}}
\newcommand{\Id}{\mbox{1\hspace{-1.05mm}l}}   

\newcommand{\la}{\lambda}
\newcommand{\bM}{\breve{M}} 
\newcommand{\M}{{\cal M}}

\newcommand{\mo}[1]{^{\mbox{\scriptsize #1}}}
\newcommand{\N}{{\cal N}}
 
\newcommand{\Pa}{{\cal P}} 
\newcommand{\bP}{\bar{P}} 

\newcommand{\s}{{\cal S}} 
\newcommand{\bs}{\bar{{\cal S}}} 
\newcommand{\bS}{\bar{S}} 
\newcommand{\T}{{\cal T}} 
\newcommand{\Tr}{\mbox{Tr}} 
\newcommand{\bu}{\bar{u}} 
\newcommand{\U}{{\cal U}} 
 
\newcommand{\vp}{\varphi} 
\newcommand{\bw}{\bar{w}}
\newcommand{\W}{{\cal W}} 

\newcommand{\SG}{S_{\rm x}} 
\newcommand{\bSG}{\bar{S}_{\rm x}} 

\begin{document}
 
\hfill {\sc HU-EP}-04/06
 
\vspace*{1cm}
 
\begin{center}
 
{\Large \bf Lattice formulation of chiral gauge theories}
 
\vspace*{0.9cm}
 
{\bf Werner Kerler}
 
\vspace*{0.3cm}
 
{\sl Institut f\"ur Physik, Humboldt-Universit\"at, D-12489 Berlin,
Germany}
 
\end{center}

\vspace*{1cm} 

\begin{abstract}
We present a general formulation of chiral gauge theories, which admits 
Dirac operators with more general spectra, reveals considerably more
possibilities for the structure of the chiral projections, and nevertheless
allows appropriate realizations. In our analyses we use two forms of the 
correlation functions which both also apply in the presence of zero modes 
and for any value of the index. To account properly for the conditions on
the bases the concept of equivalence classes of pairs of them is introduced.
The behaviors under gauge transformations and under CP transformations 
are unambiguously derived.
\end{abstract}

\section{Introduction}

For the non-perturbative definition of the quantum field theories of particle
physics only the lattice approach is available. While for QCD it is      
nowadays clear how to proceed, in the case of chiral gauge theories still
quite some details remain to be clarified. To make progress within this 
respect is the aim of the present paper. In this context generalizations
help to see which features are truly relevant.

Starting from the basic structure of chiral gauge theories on the lattice,
which has been introduced in the overlap formalism of Narayanan and Neuberger
\cite{na93} and in the formulation of L\"uscher \cite{lu98}, we have recently
worked out a generalization \cite{ke03}. Our definitions of operators there 
have referred to a basic unitary and $\ga$-Hermition operator. While this has 
provided a guide to many detailed results, to base the formulation on this 
operator introduces unnecessary restrictions. Therefore we here develop a 
more general formulation which does not rely on this operator. 

We further base our developments now on the concept of equivalence classes
of pairs of bases which appears more appropriate than our previous use of
separate classes since it exploits the respective freedom fully and also 
turns out to be more natural in view of the structures we find for the 
general chiral projections.

It appears worthwhile to emphasize that in contrast to other 
approaches we take care that our results also hold in the presence of zero 
modes of the Dirac operator and for any value of the index.

In Section 2 we start with relations for the Dirac operators, removing 
restrictions on their spectra, which have been inherent in all analytical 
forms so far. The resulting operators are seen to have still realizations 
with appropriate locality properties and methods of numerical evaluation. 
A discussion of the locations of the spectra illustrates the respective 
new possibilities. We also introduce the generalization of the unitary 
operator of previous formulations and see its connection to the index.

In Section 3 we derive the properties of the chiral projections for given
Dirac operator, revealing considerably more possibilities for their 
structures. This derivation is based on the spectral representations of
the operators and a careful consideration of details related to the Weyl 
degrees of freedom. 

We further express the chiral projections in an alternative form which is 
particularly useful for the study of CP properties as well as with respect 
to applications of generalized chiral symmetries of the Dirac operator. We 
also see that there are appropriate realizations of the more general chiral 
projections.

In Section 4 we consider correlation functions and analyze the emerging 
conditions on the bases. We first formulate fermionic functions in 
terms of alternating multilinear forms. The requirement of invariance of 
these functions imposes restrictions on possible basis transformations. 
To account for this we introduce the concept of the decomposition into 
equivalence classes of pairs of bases and discuss its crucial significance.

The relations for the chiral projections, which we find in our analysis,
imply corresponding relations for the bases. This allows us to obtain a 
form of the correlation functions which involves a determinant and separate 
zero mode terms. It has the virtue that the contributions of particular 
amplitudes become explicit also in the general case considered.

In Section 5 we give a general derivation of gauge-transformation properties
closing a loophole in our previous derivation. The cases where both of the 
chiral projections are gauge-field dependent and where one of them is 
constant are treated separately. We add a discussion of perturbation 
theory showing the necessity of the anomaly cancelation condition 
in the continuum limit.

In Section 6 we similarly give an improved derivation of CP-transformation
properties and confirm certain features for the more general structures
here, too.

In Section 7 we consider some details of interest also in terms of
gauge-field variations, which appears useful for making contact to the work
of L\"uscher \cite{lu98}. 

In Section 8 we collect some conclusions.

\section{Dirac operator}\se

\subsection{Spectral properties}

We consider a finite lattice and require the Dirac operator to be normal, 
$[D\dg,D]=0$, and $\ga$-Hermitian, $D\dg=\ga D\ga$. It then has the spectral
representation
\be
D=\sum_j\hat{\la}_j(P_j^++P_j^-) 
+\sum_{k}(\la_kP_k\mo{I}+\la_k^*P_k\mo{II}),
\label{specd}
\ee
where the eigenvalues are all different and satisfy $\mbox{Im }\hat{\la}_j=0$ 
and $\mbox{Im }\la_k>0$. For the projections the relations $\ga P_j^{\pm}=
P_j^{\pm}\ga=\pm P_j^{\pm}$ and $\ga P_k\mo{I}=P_k\mo{II}\ga$ hold. With 
this we have $\Tr(\ga P_k\mo{I})=\Tr(\ga P_k\mo{II})=0$, $\Tr\,P_k\mo{I}=
\Tr\,P_k\mo{II}=\,:N_k$ and $\Tr\,P_j^{\pm}=\,:N_j^{\pm}$.

Presence of zero modes of $D$ means that one of the $\hat{\la}_j$ is zero,
which we take to be that with $j=0$. Since the zero-mode part of $D$ commutes 
with $\ga$, the projector on the respective space is of form 
$P_0^++P_0^-$ with $P_0^{\pm}$ having the properties described for $P_j^{\pm}$
above. Accordingly the index of $D$ is given by $I=N_0^{+}-N_0^{-}$.

In terms of the introduced projections the identity operator can be 
represented by 
\be
\Id=\sum_j(P_j^++P_j^-)+\sum_k(P_k\mo{I}+P_k\mo{II}),
\label{ID}
\ee
which implies the relation
\be
\Tr(\ga\Id)=\sum_j(N_j^+-N_j^-)=0.
\label{SU}
\ee
It is to be noted that with \re{ID} and \re{SU} we also have 
\be
\sum_jN_j^++\sum_kN_k=\sum_jN_j^-+\sum_kN_k=\h\Tr\,\Id=\,:d. 
\label{IDH}
\ee

\subsection{Associated unitary operator}

In the absence of zero modes of $D$ the operator $V=-DD^{\dag-1}$ is a well
defined unitary and $\ga$-Hermitian operator. To include the case with
zero modes we require
\be
D+D\dg V=0,
\label{DDV}
\ee
in addition to unitarity and $\ga$-Hermiticity, which fixes $V$ up to the 
sign of the $P_0^++P_0^-$ term in the spectral representation. Taking the
positive one we get
\be
V=P_0^++P_0^--\sum_{j\ne0}(P_j^++P_j^-)-\sum_{k}\Big(\frac{\la_k}{\la_k^*}P_k
\mo{I}+\frac{\la_k^*}{\la_k}P_k\mo{II}\Big).
\label{specv}
\ee
With this we obtain $\Tr(\ga V)=N_0^{+}-N_0^{-}-\sum_{j\ne0}(N_j^+-N_j^-)$,
which together with \re{SU} gives for the index
\be
I=\h\Tr(\ga V),
\label{IV}
\ee
i.e.~ still the form which in Refs.~\cite{ke02,ke03} has been seen to 
generalize earlier results \cite{na93,ha98,lu98a}. The negative sign for the 
$P_0^++P_0^-$ term instead leads to an operator $\tilde{V}$ with 
$\Tr(\ga\tilde{V})=0$.

In \re{specv} the projector related to the eigenvalue $-1$ decomposes 
into projections corresponding to those associated to the different real 
eigenvalues of $D$ which occur in addition to zero. Similarly for complex 
eigenvalues $\la_k=r_k\e^{i\alpha_k}$ the associated eigenvalues of $V$ do 
not differ for $r_{k'}\ne r_k$ if $\alpha_{k'}=\alpha_k$. Furthermore 
because of $0<\alpha_k<\pi$ the factors $-\frac{\la_k}{\la_k^*}=
\e^{i(2\alpha_k-\pi)}$ of form $\e^{i\beta_k}$ for $0<\beta_k<\pi$ have
contributions of the types $\beta_{k'}=2\alpha_{k'}-\pi\,$ {\it and} 
$\,\beta_{k''}=\pi-2\alpha_{k''}$ while $\beta_k=0$ is obtained for 
$\alpha_k=\pm ir_k$. Comparing all this with the spectral representation 
\be
P_0^++P_0^--P_1^+-P_1^-+\sum_{k}\Big(\e^{\eta_k}P_k\mo{I}+\e^{-\eta_k}
P_k\mo{II}\Big),\qquad\quad0<\eta_k<\pi,
\label{specv0}
\ee
of the special case of the operator $V$ in Refs.~\cite{ke02,ke03} it becomes 
obvious that \re{specv} resulting from $D$ is within several respects a 
considerable generalization.

Since the Dirac operators in Refs.~\cite{ke02,ke03} constructed on the basis 
of \re{specv0} are ones admitting only one real eigenvalue of $D$ in 
addition to zero and, as the comparison with \re{specv} also shows, with
restrictions of the complex eigenvalues, too, we see that {\it not} to 
start from  \re{specv0} as we do here opens much more general possibilities 
for $D$.

The classes of Dirac operators with the indicated restrictions 
\cite{ke02,ke03} contain as the simplest case Ginsparg-Wilson
(GW) fermions \cite{gi82} for which $D$ is of form $D=\rho(\Id-V)$ with a 
real constant $\rho$. Further special cases are the ones proposed by 
Fujikawa \cite{fu00}, the extension of them \cite{fu02} and the various 
examples constructed in Ref.~\cite{ke02}. 

In the GW case with $D=\rho(\Id-V)$ the explicit realization of Neuberger 
\cite{ne98} creates the unitary operator $V$ by the normalization 
of another operator, namely of the Wilson-Dirac operator. This construction 
has been generalized in Ref.~\cite{ke02} and still more in Ref.~\cite{ke03} 
to apply to specific subclasses, respectively, of the general classes of
Dirac operators there.

Another type of explicit construction of $V$ in the GW case is contained in 
a definition proposed for $D$ by Chiu \cite{ch01}. With it, however, on the 
finite lattice a non-vanishing index is prevented by a sum rule \cite{ch98}, 
which is the GW special case of \re{SU}. In Ref.~\cite{ke02a} we have 
pointed out that the respective $V$ is of the Cayley-transform type and shown 
that on the finite lattice, this type generally does not allow a 
non-vanishing index, while in the continuum limit due to the unboundedness 
of $D$ it does.

\subsection{Particular realizations}

The conditions of normality and $\ga$-Hermiticity have been seen here to lead 
already to several general relations for chiral fermions. Further restrictions
result from the connection of locality and chiral properties. A definite 
requirement which can be formulated within this respect is that locality of 
$D$ should imply appropriate properties of the propagator. 

In the GW case from $\{\ga,D\}=\rho\1D\ga D$ one gets $\{\ga,D\1\}=\rho\1\ga$ 
provided that $D\1$ exists, which means that the propagator is chiral up 
to a local contact term. This can also be expressed by $D\1+D^{\dag-1}=\rho\1$. 
The generalization of the latter condition is $D\1+D^{\dag-1}=2F$ where $F$ 
is a local operator. To obtain a condition which applies also in the presence 
of zero modes of $D$ we multiply this by $D$ and $D\dg$ getting $\h(D+D\dg)=
DFD\dg=D\dg FD$, which indicates that
\be
[F,D]=0,\quad\qquad F\dg=F,\quad\qquad [\ga,F]=0.
\label{FF}
\ee
To account for this we require $F$ to be a non-singular function of $D$,
in detail a Hermitian one of the Hermitian arguments $DD\dg$ and $\h(D+D\dg)$,
and impose the condition\footnote{
  This differs from the GW relation \cite{gi82} 
  $\{\ga,D\}=2D\ga RD$ with $[R,D]\ne0$ in which $D$ is not normal. Though 
  then nevertheless no eigennilpotent spoils the subspace of zero modes 
  \cite{ke01a}, the effect in other terms remains obscure and the analysis 
  of interest here gets not feasible.}
\be
{\textstyle\h}(D+D\dg)=DD\dg\;F\big(DD\dg,{\textstyle\h}(D+D\dg)\big),
\label{DDF}
\ee
in which $F$ must be local for local $D$ (with exponential locality being
sufficient).

The operators in Ref.~\cite{fu02} correspond to the special choice of $F$ 
with the dependence on $D\dg D$ only and with a monotony requirement 
which implies restriction to only one real eigenvalue in addition 
to zero. To study the choice $F=F(DD\dg)$ without such a restriction we 
consider the special case of a polynomial $F=\sum_{\nu=0}^M\C_{\nu}
(DD\dg)^{\nu}$ with real coefficients $\C_{\nu}\,$. The eigenvalues of 
$D$ then satisfy 
\be
\mbox{Re}\,\la=\sum_{\nu=0}^M\C_{\nu}|\la|^{2(\nu+1)},
\label{LA}
\ee
which describes the location of the spectrum. From this it is obvious that 
$\la=0$ is always included and that the other real eigenvalues are subject 
to $\sum_{\nu=0}^M\C_{\nu}|\hat{\la}|^{2\nu+1}=1$, which indicates that 
in this example one can have up to $2M+1$ further real eigenvalues.

If only one of the coefficients $\C_{\nu}$ differs from zero this gives 
the proposal of Fujikawa \cite{fu00} (for $\nu=0$ the GW case) with
only one real eigenvalue in addition to zero. If only $\C_0$ and $\C_1$
are non-zero for $\C_0>0$ and $\C_1>0$ one gets an example given in 
Ref.~\cite{ke02} still with only one additional real eigenvalue. However, 
for $\C_0^3/\C_1<-27/4$ three different real eigenvalues get possible 
in addition to zero. Then the location of the spectrum is described by two 
closed curves, one through zero and further one surrounding it. 

An overview in the general case is obtained by putting $\la=r\e^{i\alpha}$
and noting that the spectral function $f$ associated to $F$ is a real 
function with the dependences $f(r,\cos\alpha)$ and that  \re{DDF} in terms 
of spectral functions reads $r\cos\alpha=rf(r,\cos\alpha)$. Obviously 
$\la=0$ is always included in this and the other values are subject to
the equation $\cos\alpha=f(r,\cos\alpha)$. 

For the more general operators $D$ here the constructions relying on 
the special case of $V$ with the spectral representation \re{specv0} are 
no longer available. Since one cannot count on the existence of explicit 
analytical forms, one has to find other methods which on the one hand side 
provide a theoretical description and on the other numerical approximations. 

The extension of the method of chirally improved fermions \cite{ga00} of 
the GW case, which is based on a systematic expansion of the Dirac operator, 
is applicable also in the case considered here. Indeed, the mapping of the 
GW equation to a system of coupled equations there can as well be done for 
the more general relation \re{DDF}. Apart from providing the theoretical 
possibility, appropriate choices in \re{DDF} could even be advantageous 
in numerical work.

\section{Chiral projections}\se

\subsection{Basic properties}

We introduce chiral projections $P_{\pm}$ and $\bP_{\pm}$ with
$P_{\pm}\dg=P_{\pm}$, $\bP_{\pm}\dg=\bP_{\pm}$ and $P_++P_-=\bP_++\bP_-=\Id$,
requiring that they satisfy
\be
\bP_{\pm}D=DP_{\mp}.
\label{DP}
\ee 
With this we get the decomposition of the Dirac operator into Weyl operators,
\be
D=\bP_+DP_-+\bP_-DP_+,
\ee
in which $\bP_{\pm}DP_{\mp}= DP_{\mp}=\bP_{\pm}D$. Furthermore, since with 
\re{DP} also $D\dg\bP_{\pm}=P_{\mp}D\dg$ holds, we obtain the relations
\be
[P_{\mp},D\dg D]=0,\quad\qquad[\bP_{\pm},DD\dg]=0.
\label{COM}
\ee

\subsection{Spectral structure}

Because with \re{specd} we have the representation 
\be
D\dg D=\sum_j\hat{\la}_j^2(P_j^++P_j^-) 
+\sum_{k}|\la_k|^2(P_k\mo{I}+P_k\mo{II}),
\label{specdd}
\ee
according to \re{COM} the chiral projections $P_-$ and $\bP_+$ decompose as
\be
P_-=\sum_jP_j^{(+,-)}+\sum_kP_k^{\rm R},\quad\qquad
\bP_+=\sum_j\bP_j^{(+,-)}+\sum_k\bP_k^{\rm R},
\label{PbP}
\ee
where $P_j^{(+,-)}$ and $\bP_j^{(+,-)}$ project within the subspace on which 
$P_j^++P_j^-$ projects, while $P_k^{\rm R}$ and $\bP_k^{\rm R}$
project within that on which $P_k\mo{I}+P_k\mo{II}$ projects. 

Noting that $P_k\mo{I}$, $P_k\mo{II}$, $\ga P_k\mo{I}$, $\ga P_k\mo{II}$ 
commute with $P_k\mo{I}+P_k\mo{II}$, imposing the general conditions $P^2=P$ 
and $P\dg=P$ and according to \re{DP} requiring $\bP_k^{\rm R}D=DP_k^{\rm R}$ 
we obtain the relations
\ba
P_k^{\rm R}=c_kP_k\mo{I}+(1-c_k)P_k\mo{II}-\sqrt{c_k(1-c_k)}\ga
(\e^{i\vp_k}P_k\mo{I}+\e^{-i\vp_k}P_k\mo{II}),\nonumber\\
\bP_k^{\rm R}=c_kP_k\mo{I}+(1-c_k)P_k\mo{II}+\sqrt{c_k(1-c_k)}
\ga\big(\e^{-i\bar{\vp}_k}P_k\mo{I}+\e^{i\bar{\vp}_k}P_k\mo{II}\big),
\label{PbP1}
\ea
with real coefficients $c_k$ satisfying $0\le c_k\le1$ and phases $\vp_k$,
$\bar{\vp}_k$ being for $c_k(1-c_k)>0$ subject to 
\be
\e^{i(\vp_k+\bar{\vp}_k-2\alpha_k)}=-1\qquad\mbox{ with }\qquad
\e^{i\alpha_k}=\la_k/|\la_k|,\qquad 0<\alpha_k<\pi,
\label{PH}
\ee
and where we have for the dimensions
\be
\Tr\,P_k^{\rm R}=\Tr\,\bP_k^{\rm R}=\Tr\,P_k\mo{I}=\Tr\,P_k\mo{II}=N_k. 
\ee

Similarly since $P_j^+$ and $P_j^-$ commute with $P_j^++P_j^-$ in accordance
with \re{DP} we arrive at
\be
P_j^{(+,-)}=\bP_j^{(+,-)}\qquad\mbox{ for }\qquad j\ne0.
\label{PJ}
\ee
For the numbers of anti-Weyl and Weyl degrees of freedom $\bar{N}=\Tr\,\bP_+$
and $N=\Tr\,P_-$ we therefore obtain 
\be
\bar{N}-N=\Tr\,\bP_0^{(+,-)}-\Tr\,P_0^{(+,-)},
\ee
which requiring $\bar{N}-N=I$ leads to
\be
\bP_0^{(+,-)}= P_0^+,\qquad P_0^{(+,-)}=P_0^-.
\label{PbP2} 
\ee

We next note that we now have 
\be
\bar{N}+N=N_0^++N_0^-+2\sum_{j\ne0}\Tr\,P_j^{(+,-)}+2\sum_kN_k\qquad
\mbox{ for }\qquad I=0,
\ee
so that in view of \re{IDH} to get $\bar{N}+N=\Tr\,\Id=2d$ for $I=0$ we must put
\be
P_j^{(+,-)}=P_j^+\quad\mbox{ for }\quad j\ne0\qquad\quad\mbox{\bf or }\quad
\qquad P_j^{(+,-)}=P_j^-\quad\mbox{ for }\quad j\ne0.
\label{PbP3}
\ee
For these choices we get in the general case
\be
\bar{N}=d,\;\;N=d-I\qquad\quad\mbox{\bf or }\qquad\quad\bar{N}=d+I,\;\;N=d,
\hspace{10mm}
\label{NbN}
\ee
respectively.

Inserting \re{PbP2} and \re{PbP3} into \re{PbP} we have
\be
P_-=P_0^-+\sum_{j\ne0}P_j^{\pm}+\sum_kP_k^{\rm R},\quad\qquad
\bP_+=P_0^++\sum_{j\ne0}P_j^{\pm}+\sum_k\bP_k^{\rm R},
\label{PbPf}
\ee
which taking the traces gives for the dimensions
\be
N=N_0^-+L,\qquad \bar{N}=N_0^++L,\qquad\quad L=\sum_{j\ne0}N_j^{\pm}+\sum_kN_k.
\label{LL}
\ee
Relation \re{LL} shows that there is a $L\times L$ submatrix $\bM$ of 
the chiral matrix $M=\bu\dg Du$ from which the zero-mode parts are removed 
and that solely the latter can make $M$ non-quadratic.

We see now that for given Dirac operator there is still freedom in the 
details of the chiral projections, which consists in the posssible two 
choices in \re{PbP3} and furthermore in that of the coefficients $c_k$ 
and the phases $\varphi_k$ and $\bar{\varphi}_k$ in \re{PbP1}. 

The index introduced in the Atiyah-Singer framework \cite{at68} on the basis 
of the Weyl operator corresponds to the one defined here for the Dirac 
operator. Since the non-zero modes there come in chiral pairs, our relation 
$\bar{N}-N=I$ has the appearance of a transcription to the finite case of 
what one has there. However, the effects we observe for $\bar{N}+N$ for 
different values of $I$ here, have no counterpart there. 
The sum rule \re{SU} for the index of $D$ reflects the fundamental 
structural difference between the two approaches \cite{ke03a}. While in 
the Atiyah-Singer case the respective effects are accommodated by the 
space structure, in lattice theory (and thus in the quantized theory it 
is to define) the space structure is independent of the index.

\subsection{Alternative form}

To see further properties of the chiral projections we express them by 
\be
P_-=\h(\Id-\ga G),\quad\qquad\bP_+=\h(\Id+\bG\ga),
\label{GaG}
\ee
which implies that $G$ and $\bG$ are unitary and $\ga$-Hermitian operators.
According to \re{DP} they satisfy
\be
D+\bG D\dg G=0.
\label{DG}
\ee
Using the relations for $P_-$ and $\bP_+$ derived before we obtain 
the spectral representations
\ba
G=P_0^++P_0^-\mp\sum_{j\ne0}\big(P_j^++P_j^-)+\sum_{k\;
(0<\phi_k<\pi)}\big(\e^{i\phi_k} P_k\mo{A}+\e^{-i\phi_k} P_k\mo{B}\big),
\nonumber\\\bG=P_0^++P_0^-\pm\sum_{j\ne0}\big(P_j^++P_j^-)+
\sum_{k\;(0<\bar{\phi}_k<\pi)}\big(\e^{i\bar{\phi}_k}\bP_k\mo{A}+\e^{-i\bar{
\phi}_k}\bP_k\mo{B}\big),\,
\label{GaG1}
\ea
in which the new quantities are related to ones introduced before by 
\ba
\cos\phi_k=a_k\cos\vp_k,\qquad\sin\phi_k=\sqrt{1-a_k^2
\cos^2\vp_k},\nonumber\\\cos\bar{\phi}_k=a_k\cos\bar{\vp}_k,
\qquad\sin\bar{\phi}_k=\sqrt{1-a_k^2\cos^2\bar{\vp}_k},\,
\label{Php}
\ea
where $a_k=2\sqrt{c_k(1-c_k)}$, 
\ba
P_k\mo{A}=\big(h_k^2P_k\mo{I}+b_k^2P_k\mo{II}-ib_kh_k\ga(P_k\mo{I}-P_k\mo{II})
\big)/(h_k^2+b_k^2),\nonumber\\P_k\mo{B}=\big(b_k^2P_k\mo{I}+h_k^2P_k\mo{II}+
ib_kh_k\ga(P_k\mo{I}-P_k\mo{II})\big)/(h_k^2+b_k^2),\nonumber\\
\bP_k\mo{A}=\big(\bar{h}_k^2P_k\mo{I}+b_k^2P_k\mo{II}-ib_k\bar{h}_k\ga
(P_k\mo{I}-P_k\mo{II})\big)/(\bar{h}_k^2+b_k^2),\nonumber\\\bP_k\mo{B}=
\big(b_k^2P_k\mo{I}+\bar{h}_k^2P_k\mo{II}+
ib_k\bar{h}_k\ga(P_k\mo{I}-P_k\mo{II})\big)/(\bar{h}_k^2+b_k^2),\,
\label{COE0}
\ea
where $b_k=1-2c_k$ and
\be
h_k=a_k\sin\vp_k+\sin\phi_k,\qquad
\bar{h}_k=a_k\sin\bar{\vp}_k+\sin\bar{\phi}_k.
\label{COE}
\ee

\hspace{0mm}From \re{GaG1} it is seen that $G=\Id$ can be obtained by choosing
the lower sign of the $j$-sums and putting $\phi_k=0$. The latter according
to \re{Php} implies that one must have $c_k=\h$ and $\vp_k=0$. Because of 
\re{PH} $\vp_k=0$ requires that $\bar{\vp}_k$ satisfies $e^{i(\bar{\vp}_k-2
\alpha_k)}=-1$. This and the opposite sign of the $j$-sum of $\bG$ 
in \re{GaG1} show that one then necessarily obtains $\bG\ne\Id$. Analogously 
in the particular case $\bG=\Id$ one finds that one gets $G\ne\Id$.

It becomes also obvious from \re{GaG1} that one has always $\bG\ne G$. This
is so because of the opposite signs of the respective $j$-sums there, which
to allow for a non-vanishing index according to \re{SU} must not vanish. 
(The $k$-sums in \re{GaG1} can be made equal by putting $\bar{\vp}_k=\vp_k$, 
in which case condition \re{PH} gets $\e^{2i(\vp_k-\alpha_k)}=-1$.)

\subsection{Special realizations}

If one puts $c_k=\h$ the operators $G$ and $\bG$ commute with $D$, as can
be seen from \re{COE0}. Then one also gets $\bG G=V$, where $V$ is the 
general operator in \re{specv}. This becomes obvious comparing \re{DDV} and 
\re{DG} and noting the sign resulting according to \re{GaG1} for the 
$P_0^++P_0^-$ term. The operators $G$ and $\bG$ then nevertheless remain 
still more general than those in Ref.~\cite{ke03}. 

The formulations of Refs.~\cite{na93,lu98} use GW fermions, 
in Ref.~\cite{na93} in the explicit form of the Neuberger operator 
\cite{ne98}. The chiral projections in these approaches in our notation 
correspond to the special choice $G=V$, $\bG=\Id$. Also in 
the GW case a generalization of this has been proposed by Hasenfratz 
\cite{ha02}, which in our notation is
\be
G=\big((1-s)\Id+sV\big)/\N,\quad\bG=\big(s\Id+(1-s)V\big)/\N,
\label{Gs}
\ee
with a real parameter $s\ne\h$ and $\N=\sqrt{\Id-2s(1-s)\big(\Id- \h(V+V\dg)
\big)}$. This is also the choice in Ref.~\cite{fu02} with the $D$ introduced
there, as is seen switching to the related $V$ which has been determined in 
Ref.~\cite{ke02}. It should be noted that for $\bG$ and $G$ satisfying \re{Gs}
one generally has $\bG G=G\bG=V$.

To obtain realizations of the more general chiral projections here the choice 
$c_k=\h$ is convenient. Then in particular the form \re{Gs} can be used 
inserting the general operators \re{specv}. Comparing  
\re{DDV} and \re{DDF} one gets the more detailed form
\be
V=\Id-2D\,F\big(DD\dg,{\textstyle\h}(D+D\dg)\big)  
\label{VV}
\ee
for this, which also has appropriate locality properties.

\subsection{Generalized chiral symmetries}

In Ref.~\cite{lu98a} L\"uscher has pointed out that in the GW case there
is a generalized chiral symmetry of the Dirac operator. The results here
give precise informations about the respective possibilities in the general
case. 

To see this in detail we note that invariance of the action
$\bar{\psi}D\psi$ under transformations $\bar{\psi}'=
\bar{\psi}\e^{i\varepsilon\bar{\Gamma}}$, $\psi'=\e^{i\varepsilon\Gamma}\psi$
with parameter $\varepsilon$ requires that the operators $\bar{\Gamma}$
and $\Gamma$ satisfy 
\be
\bar{\Gamma}D+D\Gamma=0.
\ee
Furthermore for a chiral transformations we must have
\be
\bar{\Gamma}\dg=\bar{\Gamma}=\bar{\Gamma}\1,\qquad
\Gamma\dg=\Gamma=\Gamma\1.
\ee
Now putting
\be
\bar{\Gamma}=\bG\ga,\qquad\Gamma=\ga G
\ee
we see that $\bG$ and $G$ must be unitary and $\ga$-Hermitian operators
which satisfy \re{DG}, i.e.~which are identical to the quantities $\bG$ 
and $G$ introduced before.

It thus becomes obvious that for given Dirac operator we get all the
possibilities for generalized chiral transformations described by the forms
of operators $\bG$ and $G$ derived before. It is to be emphasized in this
context that one then also generally gets $\bG\ne G$ as we have seen in
Section 3.3. We add here that one then also has $\bar{\Gamma}\ne\Gamma$, 
again because of the opposite signs of the respective $j$-sums in \re{GaG1}, 
which to allow for a non-vanishing index according to \re{SU} must not 
vanish. 

\section{Correlation functions and bases}\se

\subsection{Basic fermionic functions}

In terms of Grassmann variables non-vanishing fermionic correlation 
functions for the Weyl degrees of freedom are given by
\ba
\langle\chi_{i_{r+1}}\ldots\chi_{i_N}\bar{\chi}_{j_{r+1}}\ldots
\bar{\chi}_{j_{\bar{N}}}\rangle_{\f}=\hspace*{88mm}\nonumber\\
s_r\int\di\bar{\chi}_{\bar{N}} \ldots\di\bar{\chi}_1\di\chi_N
\ldots\di\chi_1\;\;\e^{-\bar{\chi}M\chi}\;\;\chi_{i_{r+1}}\ldots\chi_{i_N}
\bar{\chi}_{j_{r+1}} \ldots\bar{\chi}_{j_{\bar{N}}},\hspace*{6mm}
\ea
so that putting $s_r=(-1)^{rN-r(r+1)/2}$ we have
\be
\langle\chi_{i_{r+1}}\ldots\chi_{i_N}\bar{\chi}_{j_{r+1}}\ldots
\bar{\chi}_{j_{\bar{N}}}\rangle_{\f}
=\frac{1}{r!}\sum_{j_1,\ldots,j_r=1}^{\bar{N}}\,\sum_{i_1,\ldots,
i_r=1}^N\epsilon_{j_1,\ldots,j_{\bar{N}}}\epsilon_{i_1,\ldots,i_N}
M_{j_1i_1}\ldots,M_{j_ri_r}.
\label{COR0}
\ee

The fermion field variables $\bar{\psi}_{\sigma'}$ and $\psi_{\sigma}$  
are given by $\bar{\psi}=\bar{\chi}\bu\dg$ and $\psi=u\chi$ with bases 
$\bu_{\sigma'j}$ and $u_{\sigma i}$ which satisfy
\be
P_-=uu\dg,\quad u\dg u=\Id_{\rm w},\qquad\qquad\bP_+=\bu\bu\dg,\quad\bu\dg\bu=
\Id_{\rm\bw},
\label{uu}
\ee 
where $\Id_{\rm w}$ and $\Id_{\rm \bw}$ are the identity operators in the
spaces of the Weyl and anti-Weyl degrees of freedom, respectively. Now with
the fermion action $\bar{\chi}M\chi=\bar{\psi}D\psi$, in which one gets
$M=\bu\dg Du$, we obtain from \re{COR0} for fermionic correlation functions
\be
\langle\psi_{\sigma_{r+1}}\ldots\psi_{\sigma_N}\bar{\psi}_{\bar{\sigma}_{r+1}}
\ldots\bar{\psi}_{\bar{\sigma}_{\bar{N}}}\rangle_{\f}
=\frac{1}{r!}\sum_{\bar{\sigma}_1\ldots\bar{\sigma}_r}\sum_{\sigma_1,\ldots,
\sigma_r}\bar{\Upsilon}_{\bar{\sigma}_1\ldots\bar{\sigma}_{\bar{N}}}^*
\Upsilon_{\sigma_1\ldots\sigma_N}D_{\bar{\sigma}_1\sigma_1}\ldots
D_{\bar{\sigma}_r\sigma_r}
\label{COR}
\ee
with the alternating multilinear forms
\ba
\Upsilon_{\sigma_1\ldots\sigma_N}=\sum_{i_1,\ldots,i_N=1}^N\epsilon_{i_1, 
\ldots,i_N}u_{\sigma_{1}i_{1}}\ldots u_{\sigma_Ni_N},\\\bar{\Upsilon}_{
\bar{\sigma}_1\ldots{\bar{\sigma}_{\bar{N}}}}=\sum_{j_1,\ldots,
j_{\bar{N}}=1}^{\bar{N}}\epsilon_{j_1,\ldots,j_{\bar{N}}}\bar{u}_{\bar{
\sigma}_{1}j_{1}}\ldots\bar{u}_{\bar{\sigma}_{\bar{N}}j_{\bar{N}}}.
\label{FO}
\ea

General fermionic correlation functions can be constructed as linear 
combinations of the particular non-vanishing functions \re{COR}. Having the 
fermionic correlation functions, the inclusion of the gauge fields
and the definition of full correlation functions is straightforward, at
least for vanishing index $I=0$. For $I\ne0$ in Ref.~\cite{lu98} the question
of $I$-dependent complex factors multiplying the fermionic correlation 
functions has been raised. In Ref.~\cite{ha02} the importance of such factors
for the magnitude of fermion number violating processes has been stressed. 
However, there has been no theoretical principle for deciding about them. 
In Refs.~\cite{su00,fu02} it has been suggested that the modulus of them 
could possibly be one. This is supported by our observation that for the
multilinear forms in \re{COR} we have
\be
\frac{1}{N!}\sum_{\sigma_1,\ldots,\sigma_N}|\Upsilon_{\sigma_1\ldots\sigma_N}
|^2=\frac{1}{\bar{N}!}\sum_{\bar{\sigma}_1,\ldots,\bar{\sigma}_{\bar{N}}}
|\bar{\Upsilon}_{\bar{\sigma}_1\ldots\bar{\sigma}_{\bar{N}}}|^2=1,
\ee
which means that the averages of $|\Upsilon_{\sigma_1\ldots\sigma_N}|^2$ 
and of $|\bar{\Upsilon}_{\bar{\sigma}_1\ldots\bar{\sigma}_{\bar{N}}}|^2$ 
are equal to $1$ independently of the particular values of of $N$ and of 
$\bar{N}$.

\subsection{Subsets of bases}

By \re{uu} the bases are only fixed up to unitary transformations, 
$u^{(S)}=uS$, $\bu^{(\bar{S})}=\bu\bar{S}$. While the chiral
projections remain invariant under such transformations, the forms 
$\Upsilon_{\sigma_1\ldots\sigma_N}$ and $\bar{\Upsilon}_{\bar{\sigma}_1
\ldots{\bar{\sigma}_{\bar{N}}}}$ get multiplied by factors $\det_{\rm w}S$ 
and $\det_{\rm \bw}\bar{S}$, respectively. Therefore in order that general
expectations remain invariant, we have to impose
\be
{\det}_{\rm w}S\cdot{\det}_{\rm\bw}\bar{S}\dg=1.
\label{UNI}
\ee
This is so since firstly in full correlation functions only a phase factor
independent of the gauge field can be tolerated. Secondly this factor must 
be $1$ in order that in functions with more than one contribution individual
basis transformations in its parts leave the interference terms in the 
moduli of the amplitudes invariant. It should be noted that in practice 
reactions involving more that one contribution are indeed of interest.

Condition \re{UNI} has important consequences. Without it all bases related 
to a chiral projection are connected by unitary transformations. With it the 
total set of pairs of bases $u$ and $\bu$ is decomposed into inequivalent 
subsets, beyond which legitimate transformations do not connect. These subsets 
of pairs of bases obviously are equivalence classes. Because the formulation 
of the theory must be restricted to one of such classes, the question 
arises which choice is appropriate for the description of physics. 

Different ones of the indicated equivalence classes are related by pairs of 
basis transformations $S$, $\bar{S}$ for which
\be
{\det}_{\rm w}S\cdot{\det}_{\rm\bw}\bar{S}\dg=\e^{i\Theta}\qquad
\mbox{ with }\qquad\Theta\ne0
\label{NUNI}
\ee
holds.
The phase factor $\e^{i\Theta}$ then determines how the results of the 
formulation of the theory with one class differ from the results of the 
formulation with the other class.

\subsection{Relations for bases}

The relations between the chiral projections as well the as the
decompositions of them which we have found lead to corresponding properties 
of the bases. To work this out we note that with \re{specdd} and \re{PbP} 
we have 
$DD\dg\bP_k^{\rm R}=|\la_k|^2\bP_k^{\rm R}$, which using $D\dg\bP_k^{\rm R}=
P_k^{\rm R}D\dg$ (obtained from \re{DP}) becomes
\be
\bP_k^{\rm R}=|\la_k|^{-2}DP_k^{\rm R}D\dg.
\label{DPD}
\ee
Putting $P_k^{\rm R}=\sum_{l=1}^{N_k}u_l^{[k]}u_l^{[k]\dag}$ in this we see 
that 
\be
\bu_l^{[k]}=\e^{-i\Theta_k}|\la_k|\1Du_l^{[k]}
\label{BU1}
\ee
with phases $\Theta_k$ gives the representation $\bP_k^{\rm R}=\sum_{l=1}
^{N_k}\bu_l^{[k]}\bu_l^{[k]\dag}$. Furthermore, for $P_j^{\pm}
=\sum_{l=1}^{N_j^{\pm}}u_l^{\pm[j]}u_l^{\pm[j]}\,\!\dg
=\sum_{l=1}^{N_j^{\pm}}\bar{u}_l^{\pm[j]}\bar{u}_l^{\pm[j]}\,\!\dg$
with $j\ne0$ using \re{specd} we have with phases $\Theta_j^{\pm}$ 
\be
\bu_l^{\pm[j]}=\e^{-i\Theta_j^{\pm}}|\hat{\la}_j|\1Du_l^{\pm[j]}.
\label{BU2}
\ee

With \re{BU1} and \re{BU2} it becomes obvious that the $L\times L$ 
submatrix $\bM$ of the chiral matrix $M=\bu\dg Du$, from which according
to  \re{LL} the zero modes are removed, has the eigenvalues
\be
\e^{i\Theta_k}|\la_k|,\qquad\quad\e^{i\Theta_j^{\pm}}|\hat{\la}_j|,  
\label{THF}
\ee
with multiplicities $N_k$ and $N_j^{\pm}$, respectively. Its determinant
in the subspace thus is 
\be
{\det}_L\bM=\prod_{j\ne0}(\e^{i\Theta_j^{\pm}}|\hat{\la}_j|)^{N_j^{\pm}}
\prod_k(\e^{i\Theta_k}|\la_k|)^{N_k}.
\ee

The zero mode parts are described by $P_0^-=\sum_{l=L+1}^Nu_lu_l\dg$ and 
$P_0^+=\sum_{l=L+1}^{\bar{N}}\bu_l\bu_l\dg$, where the numberings with $l>L$ 
are chosen for later notational convenience. Because of $DP_0^-=DP_0^+=0$
these bases satisfy $Du_l=0$ and $D\bu_l=0$.

\subsection{Correlation functions with determinant}

Using the bases of the preceeding Subsection to work out the combinatorics 
in \re{COR0} and denoting the eigenvalues of $\bM$ by $\Lambda_l$ we 
obtain\footnote{ 
  Note that $\epsilon_{j_1,\ldots,j_r}^{i_1,\ldots,i_r}=1,-1$ or $0$ if
  $i_1,\ldots,i_r$ is an even, an odd or no permutation of $j_1,\ldots,j_r$, 
  respectively, with the special case $\epsilon_{j_1,\ldots,j_N}\equiv
  \epsilon_{j_1,\ldots,j_N}^{1,\ldots,N}$.}
\ba
\langle\chi_{i_{r+1}}\ldots\chi_{i_N}\bar{\chi}_{j_{r+1}}\ldots
\bar{\chi}_{j_{\bar{N}}}\rangle_{\f}=\nonumber\hspace{77mm}\\
{\textstyle\frac{1}{(L-r)!}}\sum_{l_{r+1},\ldots,l_L=1}^L
\Lambda_{l_{r+1}}\1\ldots\Lambda_{l_L}\1\;\epsilon_{l_{r+1}\ldots l_{L},
L+1,\ldots,\bar{N}}^{j_{r+1},\dots,j_{\bar{N}}}\;\;\epsilon_{l_{r+1}\ldots 
l_L, L+1,\ldots,N}^{i_{r+1},\dots,i_N}\;{\det}_L\bM
\label{DET0}
\ea
for $L\ge r$ (where for $L=r$ the $\Lambda$ factors and the sum are absent), 
while for $L<r$ the function vanishes. 
With this we find for the correlation functions \re{COR}
\begin{displaymath}
\langle\psi_{\sigma_{r+1}}\ldots\psi_{\sigma_N}\bar{\psi}_{\bar{\sigma}_{r+1}}
\ldots\bar{\psi}_{\bar{\sigma}_{\bar{N}}}\rangle_{\f}
=\sum_{\sigma_{r+1}',\ldots,\sigma_N'}\epsilon\,_{\sigma_{r+1}
\ldots\sigma_N}^{\sigma_{r+1}'\ldots\sigma_N'}\;\sum_{\bar{\sigma}_{r+1}',
\ldots,\bar{\sigma}_{\bar{N}}'}\epsilon\,_{\bar{\sigma}_{r+1}\ldots
\bar{\sigma}_{\bar{N}}}^{\bar{\sigma}_{r+1}'\ldots\bar{\sigma}_{\bar{N}}'}\;
{\textstyle\frac{1}{(L-r)!}}\;{\cal G}_{\sigma_{r+1}'\bar{\sigma}_{r+1}'}
\ldots
\end{displaymath}
\be
\hspace*{20mm}\ldots\,{\cal G}_{\sigma_L'\bar{\sigma}_L'}\;\;
\e^{-i\theta_{\rm z}^-}\,u_{\sigma_{L+1},L+1}\ldots u_{\sigma_NN}\;\;\;
\e^{i\theta_{\rm z}^+}\,\bu_{L+1,\bar{\sigma}_{L+1}}\dg\ldots\bu_{\bar{N}
\bar{\sigma}_{\bar{N}}}\dg\;\;{\det}_L\bM
\label{CORd}
\ee
for $L\ge r$ (where for $L=r$ the ${\cal G}$ factors,  for $L=N$ the $u$ 
factors and for $L=\bar{N}$ the $\bu$ factors are absent), while for 
$L<r$ the function vanishes. In ${\cal G}=\breve{P}_-\breve{D}\1\breve{\bP}_+
=\breve{P}_-\breve{D}\1=\breve{D}\1\breve{\bP}_+$ the operators $\breve{D}$, 
$\breve{P}_-$, $\breve{\bP}_+$ are the restrictions of $D$, $P_-$, $\bP_+$,
respectively, to the subspace on which $\Id-P_0^+-P_0^-$ projects. It should 
be noticed that for given numbers of $\psi$ and $\bar{\psi}$ fields, the 
numbers of zero modes decide which types of contributions do occur. 

The equivalence class to which the chosen pair of bases belongs is 
characterized by the value of
\be
\sum_kN_k\Theta_k+\sum_{j\ne0}N_j^{\mp}\Theta_j^{\mp}
+\theta_{\rm z}^+-\theta_{\rm z}^-,
\label{PiT}
\ee
where $\theta_{\rm z}^+$ and $\theta_{\rm z}^-$ are the phases related
to the zero modes which we have introduced to keep \re{CORd} general. The
introduction of phases suffices in this context since a general unitary 
matrix $S$ in $N$ dimensions with $\det S=\e^{i\theta}$ can be expressed by 
the product of the matrix $\e^{i\theta/N}\Id$ and of the unimodular matrix 
$S\e^{-i\theta/N}$ which is irrelevant here.

In the absence of zero modes of $D$, where $\bar{N}=N$ and $I=0$, the 
general form \re{CORd} simplifies to
\be
\langle\psi_{\sigma_{r+1}}\ldots\psi_{\sigma_N}\bar{\psi}_{\bar{\sigma}_{r+1}}
\ldots\bar{\psi}_{\bar{\sigma}_N}\rangle_{\f}
=\sum_{\sigma_{r+1}',\ldots,\sigma_N'}\epsilon\,_{\sigma_{r+1}
\ldots\sigma_N}^{\sigma_{r+1}'\ldots\sigma_N'}\;\;{\cal G}_{\sigma_{r+1}'
\bar{\sigma}_{r+1}}\ldots{\cal G}_{\sigma_N'\bar{\sigma}_N}\;\;{\det}_NM
\label{CORe}
\ee
with ${\cal G}=P_-D\1\bP_+$ and ${\det}_NM=\langle1\rangle_{\f}$.

\section{Gauge transformations}\se

\subsection{Non-constant chiral projections}

A gauge transformation $D'=\T D\T\dg$ of the Dirac operator by \re{DP}, or
by \re{DG} and \re{GaG}, implies the corresponding transformations 
\be
P_-'=\T P_-\T\dg,\qquad \bP_+'=\T\bP_+\T\dg
\label{PT}
\ee 
of the chiral projections. We first consider the case where $[\T,P_-]\ne0$
and $[\T,\bP_+]\ne0$, i.e.~where $G\ne\Id$ and $\bG\ne\Id$.  

To get the behavior of the bases it is to be noted that the conditions 
\re{uu} must be satisfied such that the relations \re{PT} hold. It is 
obvious that given a solution $u$ of the conditions \re{uu}, then $T u$ 
is a solution of the transformed conditions \re{uu}. All solutions are 
then obtained by performing basis transformations.

In addition \re{UNI} is to be satisfied, i.e.~these considerations are 
to be restricted to an equivalence class of pairs of bases. Accordingly 
the original class $uS$, $\bu\bS$ and the transformed one $u'S'$, $\bu'\bS'$ 
are related by
\be
u'S'=\T uS\s,\qquad\bu'\bS'=\T\bu\bS\bs,
\label{uS}
\ee
where $u$, $\bu$, $S$, $\bS$ satisfy \re{uu} and \re{UNI}, respectively, and 
$u'$, $\bu'$, $S'$, $\bS'$ their transformed versions. For full generality
we have also introduced the unitary transformations $\s(\T,\U)$ and 
$\bs(\T,\U)$ obeying
\be
{\det}_{\rm w}\s(\Id,\U)\big({\det}_{\rm\bw}\bs(\Id,\U)\big)^*=1,
\ee 
\ba
{\det}_{\rm w}\big(\s(\T_{\rm a},\U)\s(\T_{\rm b},\T_{\rm a}\U\T_{\rm a}\dg)
\big)\Big({\det}_{\rm\bw}\big(\bs(\T_{\rm a},\U)\bs(\T_{\rm b},\T_{\rm a}
\U\T_{\rm a}\dg)\big)\Big)^*\hspace{1.5mm}\nonumber\\={\det}_{\rm w}\s(\T_{
\rm b}\T_{\rm a},\T_{\rm b}\T_{\rm a}\U \T_{\rm a}\dg\T_{\rm b}\dg)
\Big({\det}_{\rm\bw}\bs(\T_{\rm b} \T_{\rm a}, \T_{\rm b}\T_{\rm a}\U
\T_{\rm a}\dg\T_{\rm b}\dg)\Big)^*,
\ea 
for which
\be
{\det}_{\rm w}\s\cdot{\det}_{\rm\bw}\bs\dg=\e^{i\vartheta_{\T}}
\label{NUN}
\ee
with $\vartheta_{\T}\ne0$ for $\T\ne1$ is admitted. Obviously for 
$\vartheta_{\T}\ne0$  \re{NUN} has just the form \re{NUNI} corresponding 
to the transformation to an arbitrary inequivalent subset of pairs of bases, 
so that it ultimately cannot be tolerated. Such transformations are, on the
other hand, also excluded by the covariance requirement for L\"uscher's 
current, as will be shown in Section 7. 

Inserting \re{uS} into \re{COR} we get for the correlation functions
\ba
\langle\psi_{\sigma_1'}'\ldots\psi_{\sigma_R'}'\bar{\psi}_{\bar{\sigma}_1'}'
\ldots\bar{\psi}_{\bar{\sigma}_{\bar{R}}'}'\rangle_{\f}'=\hspace{86mm}
\nonumber\\
\e^{i\vartheta_{\T}}\sum_{\sigma_1,\ldots,\sigma_R}\sum_{\bar{\sigma}_1,
\ldots,\bar{\sigma}_{ \bar{R}}}\T_{\sigma_1'\sigma_1}\ldots\T_{\sigma_R'
\sigma_R} \langle\psi_{\sigma_1}\ldots\psi_{\sigma_R}
\bar{\psi}_{\bar{\sigma}_1} \ldots\bar{\psi}_{\bar{\sigma}_{\bar{R}}}
\rangle_{\f}\,\T_{\bar{\sigma}_1\bar{\sigma}_1'}\dg\ldots
\T_{\bar{\sigma}_{\bar{R}}\bar{\sigma}_{\bar{R}}'}\dg,
\label{COV}
\ea
indicating that they transform gauge-covariantly for $\vartheta_{\T}=0$.

\subsection{One constant chiral projection}

In the special case $G\ne\Id$, $\bG=\Id$, where $\bP_+$ is constant, the 
equivalence class of pairs of bases always contains members where $\bP_+$ 
is represented by constant bases. Indeed, given the pair $u$, $\bu$, one 
can introduce a constant basis $\bu_{\rm c}$ for which $\bu=\bu_{\rm c}
\bS_{\rm y}$ holds. Then transforming $u$ as $u=u_{\rm y}S_{\rm y}$, 
where $S_{\rm y}$ is subject to ${\det}_{\rm w}S_{\rm y}={\det}_{\rm\bw}
\bS_{\rm y}$, according to \re{UNI} the pair $u_{\rm y}$, $\bu_{\rm c}$ 
is in the same equivalence class as the pair $u$, $\bu$.

For a transformed pair $u'$, $\bu'$ we analogously get the equivalent
pair $u_{\rm y}'$, $\bu_{\rm c}$. Then instead of \re{uS} we have
\be
u_{\rm y}'S'=\T u_{\rm y}S\s,\qquad \bu_{\rm c}\bS_{\rm c}={\rm const},
\label{uS1}
\ee
where $S$ and $\bS_{\rm c}$ as well as $S'$ and $\bS_{\rm c}$ satisfy 
\re{UNI} so that ${\det}_{\rm w}S'={\det}_{\rm w}S$ holds. We furthermore
note that because of $[\T,\bP_+]=0$ we can rewrite $\bu_{\rm c}$ as
\be
\bu_{\rm c}=\T\bu_{\rm c}S_{\T}
\label{uS2}
\ee
where $S_{\T}$ is unitary.  
Insertion of \re{uS1} and \re{uS2} into \re{COR} observing \re{UNI} gives 
again the form \re{COV}, however, with
\be
\e^{i\vartheta_{\T}}={\det}_{\rm w}\s\cdot{\det}_{\rm\bw}S_{\T}\dg.
\label{vS1}
\ee

The factor ${\det}_{\rm w}\s$ in \re{vS1} again corresponds to a 
transformation to an arbitrary inequivalent subset of pairs of bases, 
which ultimately is to be excluded, while 
\be
{\det}_{\rm\bw}S_{\T}\dg={\det}_{\rm\bw}(\bu_{\rm c}\dg\T\bu_{\rm c})
\label{wS1}
\ee
is a constant which we can calculate.
For the evaluation of \re{wS1} we note that with $[\T,\bP_+]=0$ and 
$\T=e^{\G}$ we get $\bu_{\rm c}\dg\T\bu_{\rm c}=\bu_{\rm c}\dg\e^{\G\bP_+}
\bu_{\rm c}$ and the simultaneous eigenequations $\G\bP_+\bu_j^{\rm d}=
\omega_j\bu_j^{\rm d}$ and $\bP_+\bu_j^{\rm d}=\bu_j^{\rm d}$. Since the 
eigenvectors $\bu^{\rm d}$ are related to the basis $\bu_{\rm c}$ by a 
unitary transformation, $\bu^{\rm d}=\bu_{\rm c}\tilde{S}$, we get 
${\det}_{\rm\bw}(\bu_{\rm c}\dg\e^{\G\bP_+}\bu_{\rm c})=\prod_j\e^{\omega_j}
=\exp(\Tr(\G\bP_+))$, so that using $P_+=\h(1+\ga)\Id$ we arrive at
\be
{\det}_{\rm\bw}\hat{S}_{\T}\dg=\exp({\textstyle\h}\Tr\,\G)
\ee 
(where in detail $\Tr\,\G=4i\sum_{n,\ell}b_n^{\ell}\,\mbox{tr}_{\g} 
T^{\ell}$  with constants $b_n^{\ell}$ and group generators $T^{\ell}$).

\subsection{Perturbation theory}

Since in continuum perturbation theory the anomaly cancelation condition
is needed to get gauge invariance of the chiral determinant, it is to be 
checked whether this holds in the continuum limit for the lattice 
approach, too. A respective analysis has been presented in Ref.~\cite{ke03} 
of which we here briefly repeat some main points.

Putting $M=M_0+M_{\rm I}$ we get on the lattice the expansion
\be
{\det}_{\rm\bw w}M=\Big(1+\sum_{\ell=1}^{\infty}z_{\ell}\Big)\;
{\det}_{\rm\bw w}M_0,
\ee
\be
z_{\ell}=\sum_{r=1}^{\ell}\frac{(-1)^{\ell+r}}{r!}\;\sum_{\rho_1=1}^{\ell-r+1}
\ldots\sum_{\rho_r=1}^{\ell-r+1}\delta_{\ell,\,\rho_1+\ldots+\rho_r}
\;\frac{t_{\rho_1}}{\rho_1}\ldots\frac{t_{\rho_r}}{\rho_r}, 
\ee
\be
t_{\rho}=\Tr\big((D_0\1\M)^{\rho}\big),\qquad\M=\bu_0M_{\rm I}u_0\dg,
\ee
with fermion loops $t_{\rho}$, free propagators $D_0\1$ and vertices $\M$. 
With $D=D_0+D_{\rm I}$, $u=u_0+u_{\rm I}$ and $\bu=\bu_0+\bu_{\rm I}$ the
vertices decompose as
\be
\M=\bP_{+0}D_{\rm I}P_{-0}+\bu_0\bu_{\rm I}\dg Du_{\rm I}u_0\dg+\bu_0
\bu_{\rm I}\dg DP_{-0}+\bP_{+0}Du_{\rm I}u_0\dg.
\label{Mpe}
\ee

In the detailed discussion of the limit the survival of terms only at
zero and at the corners of the Brillouin zone plays a central r\^ole.
It turns out that in the limit $\bP_{+0}$ and $P_{-0}$ of the
first term on the r.h.s.~of \re{Mpe} can be replaced by $\h(1+\ga)$ and 
$\h(1-\ga)$, respectively. The other terms relying on $u_{\rm I}$ 
and $\bu_{\rm I}$ are found to vanish because the related projections get 
constant.

Since in the limit the terms vanish, the compensating effect of which on 
the finite lattice provides gauge invariance of the chiral determinant,
in any case this invariance gets lost. Furthermore, then obviously also the 
particular cases with one constant chiral projection are no longer distinct. 

For the surviving contributions the agreement with usual perturbation theory
is obvious at lower order. Considering higher orders not all Dirac operators 
can provide the appropriate results, as an example in Ref.~\cite{ka78} shows.
Since the operator of this example is non-local, it can be expected that
with the locality imposed in \re{DDF} the usual expansion is reproduced to 
any order, a proof of which remains, however, to be given.

For appropriate Dirac operators in the limit arriving at the usual structure 
of the expansion, clearly the anomaly cancelation condition is needed in 
order that a gauge-invariant continuum limit can exist.

\section{CP transformations}\se

With the charge conjugation matrix\footnote{ 
  $C$ satisfies $C\gamma_{\mu}C\1=-\gamma_{\mu}^{\rm T}$ and $C^{\rm T}=-C$.
  Using Hermitian $\gamma$-matrices with $\gamma_{\mu}^{\rm T}=(-1)^{\mu}
  \gamma_{\mu}$ for $\mu=1,\ldots,4$ we choose $C=\gamma_2\gamma_4$ and get 
$\ga^{\rm T}=\ga$ and $[\ga,C]=0$ for $\ga=\gamma_1\gamma_2\gamma_3\gamma_4$.}
$C$ and with $\Pa_{n'n}=
\delta^4_{n'\tilde{n}}$, $U_{4n}^{\rm CP}=U_{4\tilde{n}}^*$ and 
$U_{kn}^{\rm CP}=U_{k,\tilde{n}-\hat{k}}^*$ for $k=1,2,3$, where 
$\tilde{n}=(-\vec{n},n_4)$, we have
\be
D(\U^{\rm CP})=\W D^{\rm T}(\U)\W\dg,\quad\W=\Pa\gamma_4C\dg,
\label{WW}
\ee
in which T denotes transposition and where $\W\dg=\W\1$. The 
behavior of $D$ by \re{DP} implies for $P_-$ and $\bP_+$ the relations 
\be
P_-^{\rm CP}(\U^{\rm CP})=\W\bP_+^{\rm T}(\U)\W\dg,\quad 
\bP_+^{\rm CP}(\U^{\rm CP})=\W P_-^{\rm T}(\U)\W\dg.
\label{PWW}
\ee
Using $I=\Tr\,\bP_+-\Tr\,P_-$ one gets $I^{\rm CP}=-I$ for the index. 

To see more details we consider the form \re{GaG},
\be
\bP_+(\U)=\h(\Id+\bG(\U)\ga),\quad\qquad P_-(\U)=\h(\Id-\ga G(\U)),
\label{PR0}
\ee
which inserted into \re{PWW} using $\{\ga,\W\}=0$ gives
\be
P_-^{\rm CP}(\U^{\rm CP})=\h\big(\Id-\ga\W\bG^{\rm T}(\U)\W\dg\big),\quad
\bP_+^{\rm CP}(\U^{\rm CP})=\h\big(\Id+\W G^{\rm T}(\U)\W\dg\ga\big).
\label{PR1}
\ee
\hspace{0mm}From \re{WW} by \re{DG} one gets $\W\bG^{\rm T}(\U)\W\dg=\bG(\U^{
\rm CP})$ and $\W G^{\rm T}(\U)\W\dg=G(\U^{\rm CP})$, so that \re{PR1} becomes 
\be
P_-^{\rm CP}(\U^{\rm CP})=\h\big(\Id-\ga\bG(\U^{\rm CP})\big),\qquad
\bP_+^{\rm CP}(\U^{\rm CP})=\h\big(\Id+G(\U^{\rm CP})\ga\big).
\label{PRC}
\ee
Obviously this differs from the untranformed relation \re{PR0} by an 
interchange of $G$ and $\bG$. Because generally $\bG\ne G$ holds, as we 
have shown in Section 3, one cannot get the symmetric situation 
of continuum theory.

In the discussion of CP properties in Ref.~\cite{ha02}, introducing the 
special form \re{Gs} in the GW case, it has been noted that this form gets
singular for $s=\h$ so that the symmetric situation cannot be obtained. 
In the investigations of CP properties in Ref.~\cite{fu02}, using the form 
\re{Gs} together with some more general $D$, a singularity has been 
encountered if a symmetric situation has been enforced. In view of our
general result that always $\bG\ne G$ this does not come as a surprise. The 
interchange of parameters under CP transformations in Ref.~\cite{fu02} 
corresponds to the interchange of $G$ and $\bG$ in the general case here. 

With the basic conditions \re{uu} and \re{UNI} being satisfied by $u$, $\bu$, 
$S$, $\bS$ as well as by $u^{\rm CP}$, $\bu^{\rm CP}$, $S^{\rm CP}$, 
$\bS^{\rm CP}$, the equivalence classes of pairs of bases transform as
\be
u^{\rm CP}S^{\rm CP}=\W\bu^*\bS^*S_{\zeta},\quad\bu^{\rm CP}
\bS^{\rm CP}=\W u^*S^*\bS_{\zeta},
\label{tCP}
\ee
where the additional unitary operators $S_{\zeta}$ and $\bS_{\zeta}$ have
been introduced for full generality. 
Inserting \re{tCP} into \re{COR} gives for the correlation functions
\ba
\langle\psi_{\sigma_1'}^{\rm CP}\ldots\psi_{\sigma_R'}^{\rm CP}\bar{\psi}_{
\bar{\sigma}_1'}^{\rm CP}\ldots\bar{\psi}_{\bar{\sigma}_{\bar{R}}'}^{\rm CP}
\rangle_{\f}^{\rm CP}=\nonumber\hspace{90mm}\\
\e^{i\vartheta_{\rm CP}}\sum_{\sigma_1,\ldots,\sigma_R}
\sum_{\bar{\sigma}_1,\ldots,\bar{\sigma}_{\bar{R}}} \W_{\bar{\sigma}_1
\bar{\sigma}_1'}\dg\ldots\W_{\bar{\sigma}_{\bar{R}}\bar{\sigma}_{\bar{R}}'}\dg
\quad\langle\psi_{\bar{\sigma}_1}\ldots\psi_{\bar{\sigma}_{\bar{R}}}
\bar{\psi}_{\sigma_1}\ldots\bar{\psi}_{\sigma_R}
\rangle_{\f}\,\W_{\sigma_1'\sigma_1}\ldots\W_{\sigma_R'\sigma_R}.
\ea
where
\be
\e^{i\vartheta_{\rm CP}}=
{\det}_{\rm\bw}S_{\zeta}\cdot{\det}_{\rm w}\bar{S}_{\zeta}\dg.
\label{cCP}
\ee
This factor is subject to the condition that repetition of the transformation
must lead back, which is satisfied by restricting $S_{\zeta}$ and 
$\bar{S}_{\zeta}$ to choices for which $\vartheta_{\rm CP}$ is a universal 
constant. Then the factor $\e^{i\vartheta_{\rm CP}}$ gets irrelevant in full 
correlation functions so that, without restricting generality, one may put 
$\vartheta_{\rm CP}=0$.

The discussion in Ref.~\cite{fu02} has been based on a generating functional 
the content of which is similar to the respective special case of \re{CORd}.
It does not account for the restrictions due to the number of 
zero modes explicit in \re{CORd}. The non-unimodular transformation 
applied to it is not appropriate \cite{ke03}. Instead of \re{tCP} a 
respective relation without the basis transformations has been used.

\section{Variational approach}\se

\subsection{General relations}

We define general gauge-field variations for a function $\phi(\U)$ by
\be
\delta\phi(\U)=\frac{\di\phi\big(\U(t)\big)}{\di t}\bigg|_{t=0}\,,\qquad 
\U_{\mu}(t)=\e^{t\G_{\mu}^{\rm left}}\U_{\mu}\e^{-t\G_{\mu}^{\rm right}},
\label{DEF}
\ee
where $(\U_{\mu})_{n'n}=U_{\mu n}\delta^4_{n',n+\hat{\mu}}$ and 
$(\G_{\mu}^{\rm left/right})_{n'n}=B_{\mu n}^{\rm left/right}\delta^4_{n',n}$.
The special case of gauge transformations is then described by
\be
\G_{\mu}^{\rm left}=\G_{\mu}^{\rm right} =\G.
\ee

To see the consequence of the general condition \re{UNI} we vary its 
logarithm which gives
\be
\Tr_{\rm w}(S\dg\delta S)-\Tr_{\rm \bw}(\bar{S}\dg\delta\bar{S})=0.
\label{vUNI}
\ee
Instead of ${\det}_{\rm w}S\cdot{\det}_{\rm\bw}
\bar{S}\dg=1$, as needed for reactions with more than one contribution,
\re{vUNI} reflects the weaker condition ${\det}_{\rm w}S
\cdot{\det}_{\rm\bw}\bar{S}\dg=$ const. Relation \re{vUNI} can also be
expressed in terms of bases as
\be
\Tr\big(\delta(uS)(uS)\dg\big)-\Tr\big(\delta(\bu\bS)(\bu\bS)\dg\big)
=\Tr(\delta u\,u\dg)-\Tr(\delta\bu\,\bu\dg),
\label{uUNI}
\ee 
which indicates that $\Tr(\delta u\,u\dg)-\Tr(\delta\bu\,\bu\dg)$ remains
invariant within the extended equivalence class of pairs of bases specified
by ${\det}_{\rm w}S\cdot{\det}_{\rm\bw}\bar{S}\dg=$ const.

Applying variations to the basic conditions \re{uu} one can derive many 
relations. All of such relations are weaker conditions than the original 
ones. A particular example, considered in Ref.~\cite{lu98}, is the relation
\be
\Tr\big(P_-[\delta_1P_-,\delta_2P_-]\big)=
\delta_1\Tr(\delta_2u\;u\dg)-\delta_2\Tr(\delta_1u\;u\dg)
+\Tr(\delta_{[2,1]}u\;u\dg),
\label{IC}
\ee
for which we have generators $\G_{\mu(1)}^{\rm left}$, $\G_{\mu(1)}^
{\rm right}$ and $\G_{\mu(2)}^{\rm left}$, $\G_{\mu(2)}^{\rm right}$ and 
$[\G_{\mu(2)}^{\rm left},\G_{\mu(1)}^{\rm left}]$, $[\G_{\mu(2)}^{\rm right},
\G_{\mu(1)}^{\rm right}]$. We emphasize that \re{IC} follows solely from 
$P_-=uu\dg$, $u\dg u=\Id_{\rm w}$ and is not subject to further restrictions.

In the special case where zero modes of $D$ are absent and where its index is 
zero one can consider the effective action and obtains for its variation
\be
\delta\ln{\det}_{\rm\bw w}M=\Tr(P_-D\1\delta D)+\Tr(\delta u\,u\dg)-
\Tr(\delta\bu\,\bu\dg).
\label{EFF}
\ee
Because of \re{uUNI} this is invariant within the respective extended
equivalence class of pairs of bases. It is to be noted that in the 
presence of zero modes no longer only variational terms of type \re{uUNI} 
occur for the bases, as is obvious from \re{CORd}.

\subsection{Gauge transformations}
 
In the special case of gauge transformations we can use the definition 
\re{DEF} and the finite transformation relations to get the related
variations. For operators with ${\cal O}(\U(t))=\T(t)\,{\cal O}(\U(0))\,
\T\dg(t)$ and $\T(t)=\e^{t\G}$ this gives 
\be
\delt{\cal O}=[\G,{\cal O}].
\ee
For the bases in the case $[\T,P_-]\ne0$, $[\T,\bP_+]\ne0$ according 
to \re{uS} we have $u(t)=\T(t)u(0)\SG(t)$, $\bu(t)=\T(t)\bu(0)\bSG(t)$ 
where $\SG=S\s S'\hspace{0mm}\dg$, $\bSG=\bS\bs\bS'\hspace{0mm}\dg$ and 
obtain
\be
\delt u=\G\,u+u\,\SG\dg\,\delt\SG,\qquad
\delt\bu=\G\,\bu+\bu\,\bSG\dg\,\delt\bSG.
\label{DU}
\ee
With these relations the terms in the effective action become 
\be
\Tr(P_-D\1\delt D)=\Tr(\G\bP_+)-\Tr(\G P_-),
\ee
\be
\Tr(\delt u\,u\dg)=\Tr(\G P_-)+\Tr_{\rm w}(\SG\dg\,\delt\SG),\quad
\Tr(\delt\bu\,\bu\dg)=\Tr(\G\bP_+)+\Tr_{\rm\bw}(\bSG\dg\,\delt\bSG),
\ee
so that using \re{vUNI} we obtain
\be
\delt\ln{\det}_{\rm\bw w}M=\Tr_{\rm w}(\s\dg\,\delt\s)-
\Tr_{\rm\bw}(\bs\dg\,\delt\bs).
\ee

In the case $[\T,P_-]\ne0$, $[\T,\bP_+]=0$ according to \re{uS1} we get 
instead of 
\re{DU}
\be
\delt u_{\rm y}=\G\,u_{\rm y}+u_{\rm y}\,\SG\dg\,\delt\SG,\qquad
\delt\bu_{\rm c}=0,
\label{DU1}
\ee
which with \re{vUNI} and $P_+=\h(1+\ga)\Id$ gives
\be
\delt\ln{\det}_{\rm\bw w}M=\Tr_{\rm w}(\s\dg\,\delt \s)+\h\Tr(\ga\G).
\label{DEFF}
\ee

Thus we have the results expected according to those for finite 
transformations. Excluding transformations $\s$ and $\bs$ which 
arbitrarily lead to inequivalent subsets of pairs of bases here means that 
the terms involving $\delt \s$ and $\delt\bs$ do not contribute. The
latter, on the other hand, will be seen below to follow also from
the covariance requirement for L\"uscher's current.

\subsection{Special case of L\"uscher}

L\"uscher \cite{lu98} considers the variation of the effective action and
assumes $\delta\bP_+=0$ and $\delta\bu=0$ so that the last term in \re{EFF}
is absent and condition \re{vUNI} reduces to $\Tr(S\dg\delta S)=0$. With
the term $\Tr(\delta u\,u\dg)$ he defines a current $j_{\mu n}$
by
\be
\Tr(\delta u\,u\dg)=-i\sum_{\mu,n}\mbox{tr}_{\g}(\eta_{\mu n}j_{\mu n}),
\qquad\delta U_{\mu n}=\eta_{\mu n}U_{\mu n},
\ee
and requires it to transform gauge-covariantly.

His generator is given by $\eta_{\mu n}=B_{\mu,n+\hat{\mu}}^{\rm left}-
U_{\mu n}B_{\mu n}^{\rm right}U_{\mu n}\dg$ in terms of our left and right
generators. We get explicitly
\be
j_{\mu n}=i(U_{\mu n}\rho_{\mu n}+\rho_{\mu n}\dg U_{\mu n}\dg),
\qquad\rho_{\mu n,\alpha'\alpha}=\sum_{j,\sigma}u_{j\sigma}\dg
\frac{\partial u_{\sigma j}\hspace*{7mm}}{\partial U_{\mu n,\alpha\alpha'}}.
\ee
The equirement of gauge-covariance 
$j_{\mu n}'=\e^{B_{n+\hat{\mu}}}j_{\mu n}\e^{-B_{n+\hat{\mu}}}$ because 
of $U_{\mu n}'=\e^{B_{n+\hat{\mu}}}U_{\mu n}\e^{-B_n}$ implies that one 
must have 
\be
\rho_{\mu n}'=\e^{B_n}\rho_{\mu n}\e^{-B_{n+\hat{\mu}}},
\ee
which with $u'=\T uS\s S'$, $\Tr_{\rm w}(S\dg\delta S)=0$ and $\Tr_{\rm w}
(S'\hspace{0mm}\dg\delta S')=0$ leads to the condition
\be
\sum_{j,k}\s_{kj}\dg\frac{\partial\s_{jk}\hspace*{7mm}}{\partial 
U_{\mu n,\alpha\alpha'}}=0.
\ee
\hspace{0mm}From this and $\s\1=\s\dg$ it follows that
\be
\Tr(\s\dg\delt\s)=0.
\label{DELT}
\ee
Thus with \re{DEFF} in the special case considered by L\"uscher one obtains
the definite result 
\be
\delt\ln{\det}_{\rm\bw w}M=\h\Tr(\ga\G),
\ee
which leaves no freedom for changing gauge-transformation properties by a 
particular construction. 

Relation \re{DELT} shows that a transformation to an inequivalent 
subset of bases is also excluded by the covariance requirement for 
L\"uscher's current. This extends to the case where both chiral 
projections are non-constant, too, since also introducing a current 
$\bar{j}_{\mu n}$ related to $\bu$ the covariance of $j_{\mu n}-
\bar{j}_{\mu n}$ leads to $\Tr(\s\dg\delt\s)-\Tr(\bs\dg\delt\bs)=0$.

\section{Conclusions}

To make progress with the non-perturbative definition of quantized chiral 
gauge theories we have generalized previous formulations and investigated
which features are truly relevant and which properties are indeed there.

Starting with relations for the Dirac operators we have removed the 
restriction to one real eigenvalue in addition to zero and similar 
restrictions on the complex eigenvalues, which have been inherent in all 
analytical forms so far. A discussion of the locations of the spectra 
has illustrated the respective new possibilities. The generalization of 
the unitary and $\ga$-Hermition operator of previous formulations has
turned out to be again connected to the index.

The more general Dirac operators have been seen to have still realizations
with appropriate locality properties. For their numerical evaluation an
extension of the method of chirally improved fermions is suitable. The
additional freedom of these operators could possibly be advantageous for
numerical work.

We have derived the properties of the chiral projections for given Dirac 
operator using the spectral representations of the operators and carefully
considering the requirements related to the Weyl degrees of freedom. It
has turned out that there are considerable possibilities for their
structure. Nevertheless generally definite relations between Weyl and
anti-Weyl projections and a decomposition into subspaces revealing the
special r\^ole of zero modes have been found.

Expressing the chiral projections in an alternative form it has become
obvious that the symmetry between the Weyl and anti-Weyl cases
known in continuum theory can generally not be there. This, in particular,
affects the behavior under CP transformations. Using the alternative form
it has also been seen that there are appropriate realizations of the more 
general chiral projections.

We have further pointed out that the operators occurring in the alternative 
form of the chiral projections on the other hand provide generalized chiral
symmetries of the Dirac operator. Thus there is generally a whole family 
of such symmetries. Furthermore accordingly the respective left and right 
transformations are generally different. 

We have considered fermionic correlation functions in terms of alternating 
multilinear forms in order that our results also apply in the presence 
of zero modes of the Dirac operator and for any value of the index.
The requirement of invariance of these functions imposes restrictions on 
possible basis transformations. To account properly for this we have 
introduced the concept of the decomposition into equivalence classes of 
pairs of Weyl and anti-Weyl bases.

The indicated concept of pairs not only exploits the respective freedom
fully but is also natural in view of the relations between the Weyl and 
anti-Weyl projections which we find in our analysis and which imply 
corresponding relations for the bases. We have stressed that to describe 
physics one of the equivalence classes of pairs of bases is to be
chosen and that the questions arises which choice is appropriate. 

The relations between the Weyl and anti-Weyl bases together with the
decomposition into subspaces we have found has allowed us to obtain a 
further form of the correlation functions which applies also in the 
presence of zero modes and for any value of the index. It involves a 
determinant and separate zero mode terms and has the virtue that the 
contributions of particular amplitudes become explicit also in the 
general case considered. 

We have given a completely unambiguous derivation of the gauge-transformation 
properties of the correlation functions. In the case where both of the chiral
projections are gauge-field dependent the exclusion of switching to an 
arbitrary inequivalent subset of pairs of bases leads to gauge covariance. 
In the cases where one of the chiral projections is constant a factor 
depending on the particular gauge transformation remains. 

It has been noted that switching to an arbitrary inequivalent subset is also
excluded by the covariance requirement for L\"uscher's current. Thus 
obviously gauge-transformation properties on the finite lattice are fully 
determined. It has been emphasized that in the continuum limit one 
nevertheless arrives at the usual situation where the anomaly cancelation
condition is needed for gauge invariance of the chiral determinant, pointing 
out that in the limit the compensating effects of the bases are no longer 
there.

We have similarly given an unambigous derivation of CP-transformation
properties of the correlations functions. It has been seen that also for 
the more general chiral projections one cannot get the symmetric situation 
with respect to CP transformations known in continuum theory.

Finally we have considered some issues of interest also in terms of
gauge-field variations. After making certain relations precise we have 
turned to the variation of the effective action. We have then shown 
that requiring gauge covariance of L\"uscher's current prevents switching 
to inequivalent subsets of pairs of bases. Since thus there is no 
freedom for changing gauge-transformation properties by a particular
construction, the respective efforts in literature actually cannot work.

\section*{Acknowledgement}

I wish to thank Michael M\"uller-Preussker and his group for their kind
hospitality.

\end{document}